\documentclass[useAMS]{mn2e}
\usepackage{epsfig}
\usepackage{graphicx}

\newcommand{\kms}{km\,s$^{-1}$}

\def\figdir{./}
\begin{document}

\title[Spectroscopic simulations of superhumps]
  {Simulations of spectral lines from an eccentric precessing accretion disc}
\author[S. B. Foulkes et al.]
  {Stephen B. Foulkes$^{1}$, Carole A. Haswell$^{1}$, James R. Murray$^{2,3}$, Daniel J. Rolfe $^{1,2}$\\
$^1$Department of Physics \& Astronomy, The Open University, Walton Hall, Milton Keynes, MK7 6AA, UK. \\
$^2$Department of Physics \& Astronomy, University of Leicester, University Road, Leicester, LE1 7RH, UK.\\
$^3$Department of Astrophysics \& Supercomputing, Swinbourne University of Technology, Hawthorn, VIC 3122, Australia.\\
Email SBF: sbfoulkes@qinetiq.com, CAH: C.A.Haswell@open.ac.uk, JRM: jmurray@astro.swin.edu.au, DJR: dj@astro.le.ac.uk }
\date{Accepted. Received}
\pagerange{\pageref{firstpage}--\pageref{lastpage}}
\pubyear{2004}

\maketitle \label{firstpage} 
\begin{abstract}
Two dimensional SPH simulations of a
precessing accretion disc  in a  $q=0.1$ binary system (such as XTE \thinspace J1118+480)
reveal  complex and continuously varying shape, kinematics, and dissipation. The
stream-disc impact region and disc spiral density waves are prominent sources of energy
dissipation.  The dissipated energy is modulated on the period $P_{sh} =  ( {P_{orb}}^{-1}
- {P_{prec}}^{-1} )^{-1}$ with which the orientation of the disc relative to the mass
donor repeats. This superhump modulation in dissipation energy has a variation in
amplitude of ${\sim}$${10\%}$ relative to the total dissipation energy and evolves,
repeating exactly only after a full disc precession cycle.  A sharp component in the light
curve is associated with centrifugally expelled material falling back and impacting the
disc. Synthetic trailed spectrograms reveal two distinct `S-wave' features, produced
respectively by the stream gas and the disc gas at the stream-disc impact shock. These
S-waves are non-sinusoidal, and evolve with disc precession phase.  We identify the spiral
density wave emission in the trailed spectrogram.  Instantaneous Doppler maps show how the
stream impact moves in velocity space during an orbit. In our maximum entropy  Doppler
tomogram  the stream impact region emission is distorted, and the spiral density wave
emission is suppressed. A significant radial velocity modulation of the whole line
profile  occurs on the disc precession  period.  We compare our SPH simulation with a
simple 3D model:  the former is appropriate for comparison with emission lines while the
latter  is preferable for skewed absorption lines from precessing discs. 
\end{abstract}

\begin{keywords}
  accretion: accretion discs -
  X-rays: binaries -
  binaries: low mass X-ray binaries,  cataclysmic variables - 
  stars: individual: A\,0620-00, XTE \thinspace J1118+480, GRO \thinspace J1655-40 -
  methods: numerical 
\end{keywords}

\section{Introduction}

Superhumps are photometric modulations in the observed light curves found in some short
period cataclysmic variables (CVs) \cite{Patterson:1998}, and in some of
the transient low mass X-ray binaries known as soft X-ray
transients (SXTs) \cite{ZuritaEt:2002}.  In general the superhump period,
$P_{sh}$, is a few percent longer than the binary orbital period, $P_{orb}$, and
is attributed to the prograde apsidal precession of a non-axisymmetric accretion
disc. Such precession is expected in extreme mass ratio systems ($q \la 0.3$,
where $q = M_{2}/M_{1}$ is the mass ratio) in
which the disc encompasses the 3:1 resonance \cite{Lubow:1991}.  If the disc
extends to the 3:1 resonant radius, it becomes eccentric and non-axisymmetric
\cite{Murray:1998} and it will then precess under the tidal influence of the
secondary star.  In CVs, superhumps in the light curve are understood as being
powered directly from the modulated tidal stresses \cite{Osaki:1985,Murray:1998}.
This modulation is accompanied by a change in the solid angle subtended by the
disc at the central X-ray source, and it is this shape change which leads to the
irradiation-powered superhumps in  X-ray bright SXTs \cite{HaswellEt:2000}.

Material from the donor star leaves the first Lagrangian point ($L_{1}$) and flows along a ballistic trajectory towards the central primary
object. When this gas stream encounters the accretion disc material, large amounts of energy are dissipated in a brightspot
region on the outer edge of the disc. As the  eccentric accretion disc precesses in the inertial frame,  and the orbital
motion of the binary proceeds, the energy dissipated at the brightspot will be modulated as described by Rolfe, Haswell \&
Patterson (2001 hereafter RHP2001).

Spectroscopic observations of superhumping CVs \cite{Vogt:1982,HessmanEt:1992,PattersonEt:1993} and SXTs (in particular XTE\thinspace J1118+480) 
have shown that trailed spectrograms of such objects are  complex and change from night to night 
\cite{TorresEt:2001,ZuritaEt:2002,HaswellEt:2004}.  Over intervals of days or months (many times
the orbital period) the spectral line profiles evolve  from almost symmetric to enhanced
red-shifted or blue-shifted.  These changes have been attributed to a precessing accretion disc,
and herein we present  calculations which strongly support this interpretation.

In this paper we describe the results from two simulations of a precessing eccentric accretion disc. The first simulation consisted of a simple analytical three-dimensional representation of the disc and brightspot region based on RHP2001. The second model was a high-resolution smoothed particle hydrodynamics (SPH) simulation. SPH is a Lagrangian numerical method for modelling fluid flow using a set of moving particles. The fluid properties at any point within the flow field are determined by interpolating from the local particle properties. This interpolation takes the form of a summation over the 
local particles,  weighted according to their distance from the evaluation point. For a comprehensive review of the method see Monaghan (1992).

In section \ref{sec:Simulation} we 
detail the binary system modelled and the two numerical methods 
employed to model the system. Section \ref{sec:Results} presents 
our results and Section \ref{sec:Conclusions} is
devoted to discussion and conclusions.

\section{Simulations} 
\label{sec:Simulation}  
\subsection{Binary Parameters}

To simulate a typical SXT, we adopted a mass ratio $q=0.1$. 
The orbital period, $P_{orb}$, is
${\left ({ 4\pi^2 a^3/GM_t} \right )}^{1/2}$, 
where $a$ is the orbital separation,
and $M_{t}$ is the total mass of the binary system.
The mass transfer rate was $1.46 \times 10^{-9}M_{t}yr^{-1}$.

\subsection{Simple Analytical Approximation to a Precessing Eccentric Disc}

Here we briefly describe a simple model for the eccentric disc and
stream-disc impact developed from that used to model late superhumps
in the SU\,UMa CV IY\,Uma, see RHP2001. The model assumes the brightspot emission is directly proportional to the kinetic energy of the relative motion of the stream and disc flows at the impact point, i.e. the rate of energy dissipation at the brightspot is:
$$\dot{E}=\frac{1}{2}\dot{M}V_{rel\perp}^2 +
\frac{1}{2}\dot{M}V_{rel\parallel}^2\frac{V_{rel\parallel}}{\left|V_{rel\parallel}\right|}$$
where $\dot{M}$ is the mass flow rate of the stream and $V_{rel\perp}$
and $V_{rel\parallel}$ are the relative velocity of the stream and
disc resolved perpendicular and parallel to the disc respectively. The
disc velocity at radius $r$ from the primary object is assumed to be that
of an elliptical orbit around the primary object, that is:
$$|\vec{V}_{disc}|=\sqrt{GM_{1}\left(\frac{2}{r}-\frac{1}{a_{orbit}}\right)},$$
where $a_{orbit}$ is the semi-major axis of the orbit and $M_{1}$ is the mass of the primarily object. The velocity of the stream $\vec{V}_{stream}$ is simply that of the ballistic trajectory. 

The model includes a simple 3D structure for the brightspot region. An
elliptical cross-section in the r-z plane is assumed, with size
$r_{spot}$ and $h_{spot}$\footnote{See Figure 8 of RHP2001.} centred on the outer disc 
edge. $r_{spot}$,
$h_{spot}$ and the spot brightness decrease downstream from the
initial impact as $e^{-\theta^2/\Delta\theta_{spot}^2}$ where $\theta$
is the angular distance downstream from the initial impact point.
Upstream of the impact the brightspot surface is rounded off with a
hemisphere of uniform brightness equal to that at the initial impact.
We used $r_{spot}=0.03a$ and $h_{spot}=0.012a$ at $\theta=0$ ($a$ is
the binary separation).  The angular extent of the brightspot region is
set by $\Delta\theta_{spot}=\Delta\theta_0 a_{disc}/r_0$, where $r_0$
is the disc radius at the impact point and $a_{disc}$ is the
semi-major axis of the disc. This keeps the arc length of the brightspot
region roughly constant at $\Delta\theta_0 a_{disc}$ as the eccentric
disc precesses.  We used $\Delta\theta_0=30^\circ$.

The disc was modelled as an ellipse with $a_{disc}=0.31a$ 
and eccentricity 0.093 at the outer edge\footnote{The eccentricity was chose to match that of our SPH simulation}. Streamlines and brightness
contours within the disc are ellipses with size and eccentricity
decreasing smoothly to zero at the primary object, with the velocity
given by $|\vec{V}_{disc}|$. The inner boundary is near-circular with
radius 0.025$a$. The disc brightness varies as $a_{orbit}^{-1.2}$. The
disc is flared with total thickness increasing linearly from 0 at the
primary object to 0.02$a$ at the outer edge. An orbital inclination of
70$^\circ$ was used to avoid eclipses.

In the brightspot region, where the stream and disc flows merge, we 
adopted a velocity distribution 
which was non-zero 
between the
local disc velocity at that point, $\vec{V}_{disc}$, and the velocity
$\vec{V}_{disc}+(\vec{V}_{stream,0}-\vec{V}_{disc})e^{-\theta^2/\Delta\theta_{spot,v}^2}$.
Thus, at the impact point, velocities between $\vec{V}_{disc}$ and 
the ballistic stream velocity at the impact, $\vec{V}_{stream,0}$ 
contribute to the Doppler broadening of the local line profile. 
Downstream the local
velocity smoothly approaches the disc velocity. 
Upstream of the impact the exponential
factor was set to 1. $\Delta\theta_{spot,v}$ behaves as
$\Delta\theta_{spot}$ as the eccentric disc precesses. We used
$\Delta\theta_{0,v}=5^\circ$.

The local linewidths were assumed to have Gaussian profiles, with the
disc width set to the disc thermal velocity. The brightspot profile width
was the quadratic sum of the disc thermal velocity and the thermal
velocity of a 10$^4$K source decaying as
$e^{-\theta^2/4\Delta\theta_{spot}^2}$ downstream.  The decay
was imposed for consistency with the brightspot brightness, assuming brightspot brightness
$\propto T_{brightspot}^4$.

\subsection{SPH Simulation} 

\begin{figure*}
  \psfig{file=\figdir/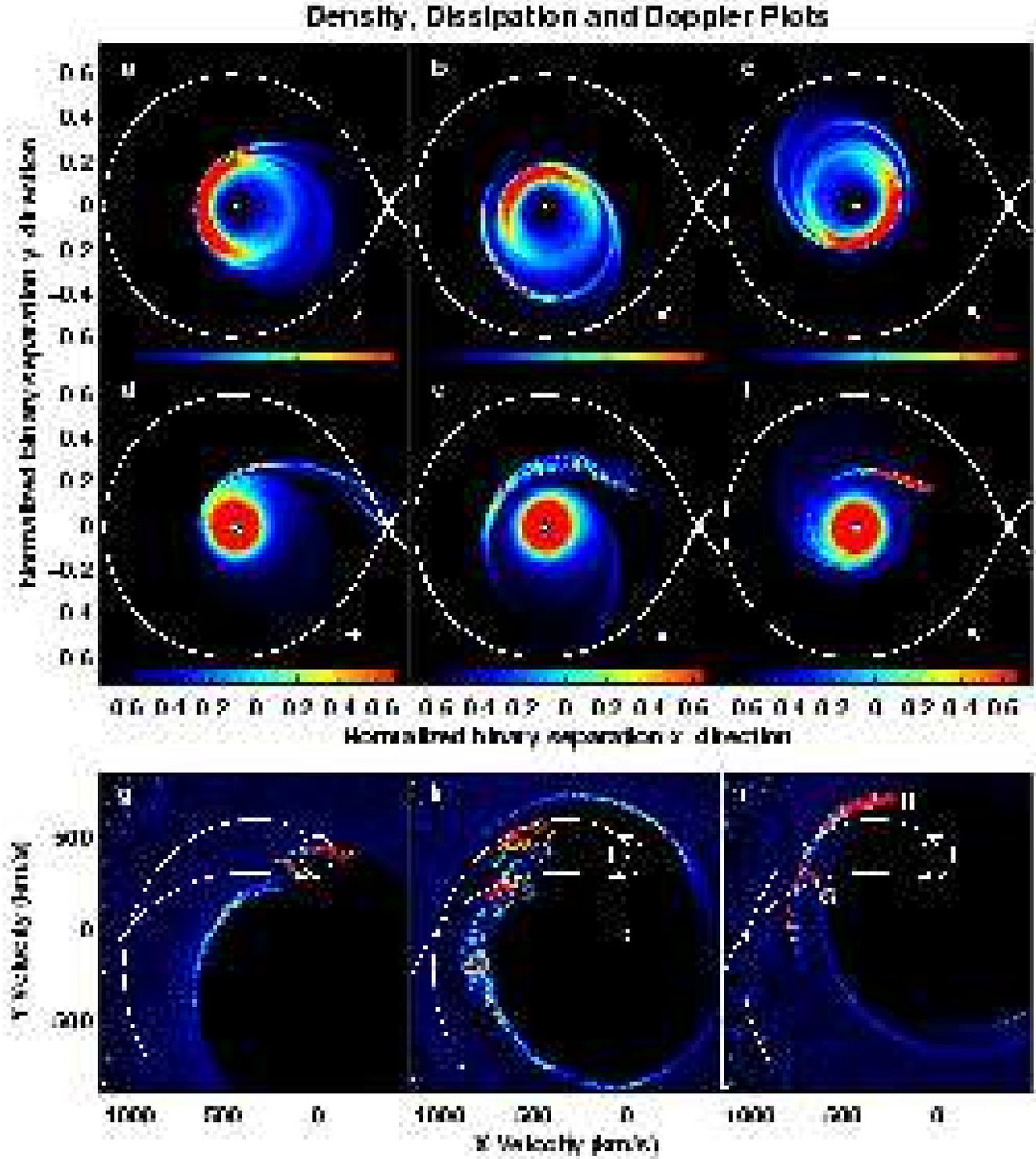,width=1.0\textwidth,angle=0}
  \caption{   
	   The top row (a, b, c) are accretion disc surface density maps using a
logarithmic colourscale for binary phases 0.00, 0.36, and 0.64 respectively. The
secondary orbits anti-clockwise with respect to the inertial frame, with mass
being added from the $L_{1}$ point at the right of each map. The solid curve is the Roche
lobe of the primary, plotted in a frame that co-rotates with the binary. The arrow
in the lower right-hand corner of each map indicates a fixed direction to an
observer. The middle row (d, e, f) are the corresponding dissipation maps plotted
with a logarithmic colourscale. The bottom row (g, h, i) are the matching Doppler
maps plotted using a linear colourscale. The letter $``G"$ on map i, indicates the
point at which the gas stream impacts on the outer edge of the accretion disc and
the letter $``H"$ indicates the arc of emission produced by the disc particles in
the impact region. 
         }
  \label{figure:density_maps}
\end{figure*}

The accretion disc was also modelled using a two-dimensional SPH computer code
that has been described in detail in Murray (1996, 1998). Modifications have been
made to convert the code from FORTRAN to C++. C++ is an Object Orientated
computer language allowing each particle to have its own private attributes (e.g.
position, velocity) and particle interactions can be coded in a more flexible
manner. The code was also extended to allow its execution on an array of
processors using the Message-Passing Interface
Standard\footnote{http://www-unix.mcs.anl.gov/mpi/}.  The SPH results presented
here were generated on a small private Linux network. 

In the SPH simulation the total system mass, $M_{t}$, and the binary separation,
$a$, were both scaled to unity. The binary orbital period, $P_{orb}$, was scaled
to $2\pi$. The accretion disc had an open inner boundary condition in the form of
a hole of radius $r_{1}=0.025a$ centred on the position of the primary object.
Particles entering the hole were removed from the simulation. Particles that
re-entered the secondary Roche lobe were also removed from the simulation as were
particles that were ejected from the disc and had a distance $> 0.9a$ from the
centre of mass of the primary. 

The artificial viscosity term of Murray (1996) was used to  improve the effective
shear viscosity in the SPH equations of motion.  In the inner regions
(${<\sim}$$0.12a$) of the accretion disc the particles  generally executed circular
orbits, and the radial shear term dominated the inner disc dissipation.  However,
in the outer disc the tidal interaction of the secondary was stronger and the bulk
term was more significant.  The disc energy dissipation was calculated using the
SPH energy equation as in  Murray (1996).  The Shakura-Sunyaev (1973) viscosity
parameters were set to  $\alpha_{low}$ of 0.1 and $\alpha_{high}$ of 1.0, and the
viscosity switched from low to high  as described in  Truss et al.
(2000). The SPH smoothing length, $h$, was  allowed to vary
in both space and time and had a maximum value of $0.01a$. 

The SPH code was started with no particles,  i.e. zero mass in the accretion
disc.  A single particle was injected into the simulation every
$0.01\Omega_{orb}^{-1}$ at the first Lagrangian point.  The sound speed in the
donor atmosphere was used as the initial speed of each particle and they followed
a ballistic trajectory. Within twenty binary orbits the resulting disc became
eccentric and encountered the Lindblad 3:1 resonance.   After the disc had reached
a mass equilibrium state, we followed its  evolution through four disc precession
cycles ($\sim$$180$ binary orbits)  to produce the results described here.  There
were approximately fifty-three thousand particles in the disc, and the average
number of `neighbours', that is the average number of particles used in the SPH
update equations, was 8.9 particles.  

The complicated hydrodynamics of the stream-disc impact have been investigated
analytically by Hessman (1999) and numerically by Armitage \& Livio, first with
the SPH technique (1996) and later with the grid based ZEUS code (1998). In the
latter paper, they found the character of the stream-overflow in particular
depended strongly upon the equation of state. Kunze, Speith \& Hessman (2001)
comprehensively studied the impact with much higher resolution SPH simulations,
for a range of binary mass ratios. They showed that a substantial fraction
(perhaps more than half)  of the stream overflowed the outer disc edge, to
impact the disc close to the circularisation radius. The latter paper is 
particularly relevent as it demonstrates that the detailed shock structure of
the impact region can be captured using the SPH technique, if sufficient
numbers of particles are deployed. The computational cost is however so great
as to make a calculation following the evolution of a disc prohibitively
expensive. 

As our SPH calculations are two dimensional, stream-overflow cannot occur and the 
brightness of the hot spot will be overemphasised by a factor  determined by cooling
in the stream. The SPH results presented in the following section will show that the
complex stream-disc interaction region dominates the energy dissipation signature of
the accretion disc. In a three dimensional SPH code stream-overflow will occur and hence
the hot spot region will have a reduced contribution to the energy dissipated.
However, the hot spot region will still contribute a significant amount to the total
dissipation observed from  the accretion disc.

\section{Results}
\label{sec:Results}  
\subsection{Disc morphology: orbital modulation}

In this subsection we follow the evolution of the SPH simulated  accretion disc through a
complete binary orbit.  The images ($a$, $b$, $c$) in Figure \ref{figure:density_maps} are
a set of surface density maps, plotted in the co-rotating frame, for orbital phases 0.00,
0.36 and 0.64 respectively. Each map uses the same logarithmic colourscale, with blue
representing low density and red high, we have also plotted the primary Roche lobe. Since
we can view the disc from any angle, we have arbitrarily fixed the conjunction of the
primary (and hence zero orbital phase) to correspond to Figure
\ref{figure:density_maps}(a).  The arrow in the lower right-hand corner of each map
indicates a fixed  direction in the inertial frame and we calculate radial velocities as they
would be observed by an observer in this direction.

These density maps show the disc to be asymmetric, eccentric  and continuously
changing shape under the tidal influence of the donor star;  its overall position
is moving only very slowly in the inertial frame. There are wrapped spiral density
waves that extend from the outermost regions of the disc down to approximately the circularisation radius of the system. These produce shear and  dissipation in the outer regions of
the disc. The spiral  density waves propagate angular momentum outwards, thus
allowing the  disc gas to move inward towards the central primary object.  The
precession of the disc means that the density maps  do not repeat from one orbit
to another; instead the maps  repeat on a period a few percent longer than the
orbital period. 

We have placed a Microsoft AVI movie of the change in surface density as function
of orbital phase on an Internet web-site (see http://physics.open.ac.uk/FHMR/ for
more information).

\subsection{Light curve}

In Figure \ref{figure:density_maps}($d$, $e$, $f$), we have plotted  viscous dissipation maps
corresponding to the density maps in Figure \ref{figure:density_maps}($a$, $b$, $c$). Each
dissipation map is represented using the same logarithmic  colourscale with blue low
dissipation and red high. It is clear from this set of images that the inner disc is in a
state of constant high dissipation.  The spiral compression waves and the impact of the gas
stream on the outer edge of the disc also generate high levels of dissipation.

We are unable to directly resolve the morphology of the accretion disc in observed interacting binary star systems. 
The simplest directly observable quantity which reveals some of the 
underlying properties is the light curve. 
Hence we use the SPH model to simulate the disc luminosity 
and generate an artificial light curve. We assume that the luminosity 
is generated through viscous dissipation
which is promptly radiated away from the point at which it was generated. 
The simulated light curve can be directly compared with the observations. We must, however, bear in mind that in observed systems there will generally be an orbital modulation in the visibility of each location in the co-rotating frame. 
Hence, our simulated light curves, in which we have essentially assumed $100\%$ visibility for each location throughout the orbit, are likely to be simpler than observed light curves.

\begin{figure*}
  \psfig{file=\figdir/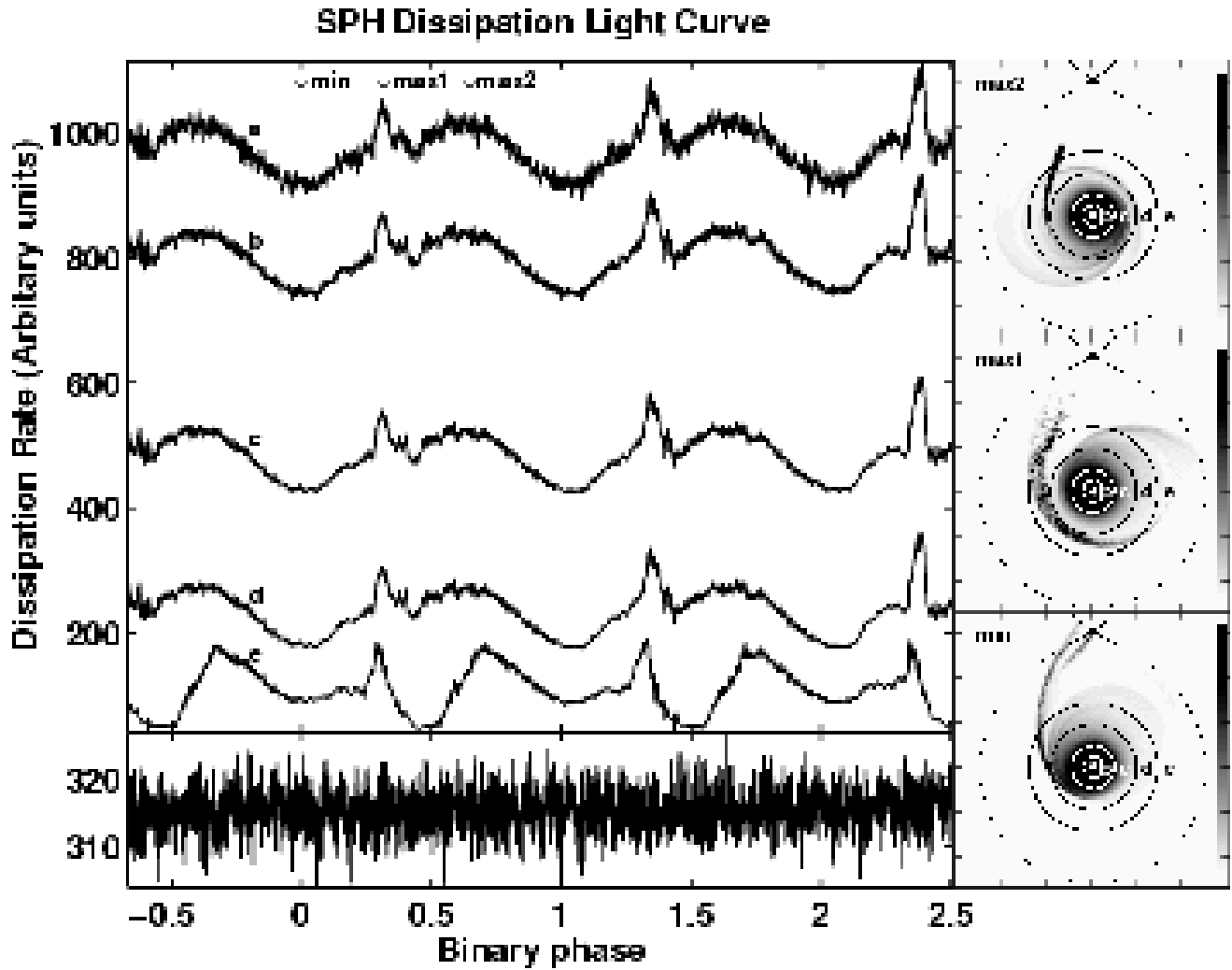,width=1\textwidth,angle=0}
   %\centerline{\psfig{file=\figdir/sph_light_curve.eps,width=\columnwidth}}
 \caption{
            The top left plot is SPH viscous dissipation light curves for different regions of the accretion disc. The top curve, labelled $a$, is for the whole accretion disc, then each curve in descending order corresponds to dissipation from the disc at radii $> 0.05a (b)$, $> 0.1a (c)$, $> 0.2a (d)$ and $> 0.3a (e)$ respectively. The light curve minimum and two maximum points are indicated. The lower left-hand plot is the signal from the disc inner region and was generated by subtracting light curve $b$ from light curve $a$. The plots on the right-hand side are disc dissipation maps that correspond to the light curve minimum and two maxima. The circles labelled $b$, $c$, $d$ and $e$ show the distances from the primarily that correspond to the light curves $b$, $c$, $d$, $e$ respectively.
          }
  \label{figure:dissipation_plot}
\end{figure*}

The upper left plot of Figure \ref{figure:dissipation_plot} shows 
five simulated light curves.
The upper curve corresponds to the total viscous dissipation
from the whole disc; the lower four curves correspond to the
four smaller annuli defined in Figure \ref{figure:dissipation_plot} caption.
The dissipation in the inner disc ($\la 0.12a$, intermediate between
`c' and `d' in Figure~\ref{figure:dissipation_plot}) 
is constant, except for stochastic noise. 
The lower left-hand plot is the dissipation from 
the innermost disc region and was generated by subtracting dissipation 
light curve $b$ from dissipation light curve $a$. Subtracting curve $c$ from $a$ and subtracting curve $c$ from $b$ would also produce similar constant dissipation curves.
Superimposed on this,
for disc regions ($r > 0.2a$)  
there is a repeating series of ``humps.'' 
These humps recur on the superhump period, $P_{sh}$. 
The modulation consists of two
separate components. The first is the relatively smooth continuous
modulation, which has a minimum at orbital phase 0.0 
in Figure \ref{figure:dissipation_plot} (corresponding to 
Figure \ref{figure:density_maps}($a, d$)) and reaches a maximum
at orbital phase $\sim$$0.64$, (corresponding to
Figure \ref{figure:density_maps}($c, f$)).
The second is the much sharper
spiky signal with a short duration 
which appears at orbital phase $\sim$$0.36$ and corresponds to
Figure \ref{figure:density_maps}($b, e$).
Photometric superhumps are invariably detected in
optical light.  Since the hot inner regions of a viscously-heated
disc contribute
relatively little to the total optical
light, 
the comparison between observation and
our simulation might best be made by comparing
one of the lower curves in Figure~\ref{figure:dissipation_plot}.

Figure \ref{figure:density_maps}(d) corresponds to the light curve minimum. 
From this map it is clear that the dissipation at
the stream-disc impact is at a minimum. 
This is because
the spiral compression arm in the outer disc
has moved such that gas is escaping from the primary 
Roche lobe and returning to the secondary. The gas leaving the $L_{1}$ 
point from the secondary encounters the disc gas instantly and 
the two flows have a low relative
velocity. 
Hence relatively little dissipation is generated by the converging
flows. At this phase the dissipation is dominated by the shear flow
in the inner regions of the disc.

As the binary orbit proceeds, a large void opens up between the $L_{1}$ point and the accretion disc, as shown in
Figure
\ref{figure:density_maps} ($b$) and ($e$)
and the spiral arm which previously extended towards the
$L_{1}$ region becomes detached from the disc under centrifugal forces 
and then falls back towards the disc. When this gas hits the disc, the impact
causes a sudden rise in dissipation,
corresponding to the spiky peak in the light curve at orbital phase $\sim$$0.36$,
and to the upper inhomogeneous arc on 
Figure~\ref{figure:density_maps}(e). 
Once this gas has been re-assimilated into the disc flow 
the dissipation falls again.  
The secondary continues its orbit around the disc, and
the stream flow and the disc flow at the stream-disc impact point
become increasingly misaligned.
As the relative velocity of the two flows increases, the dissipation
in the brightspot region correspondingly increases, until it reaches
a maximum at orbital phase $\sim$$0.64$, shown in
Figure
\ref{figure:density_maps} ($c$) and ($f$).
As the orbit proceeds, the angle between the stream and disc flows at the impact point decreases again, until the next minimum is reached 
at orbital phase 1.022. 
Similar behaviour is seen in the simple 3D model, and was fully explored
in RHP2001.
We have placed Microsoft AVI movies showing the surface dissipation
in the SPH simulation and in the analytical model 
on the Internet site http://physics.open.ac.uk/FHMR/ .

\begin{figure}
  \psfig{file=\figdir/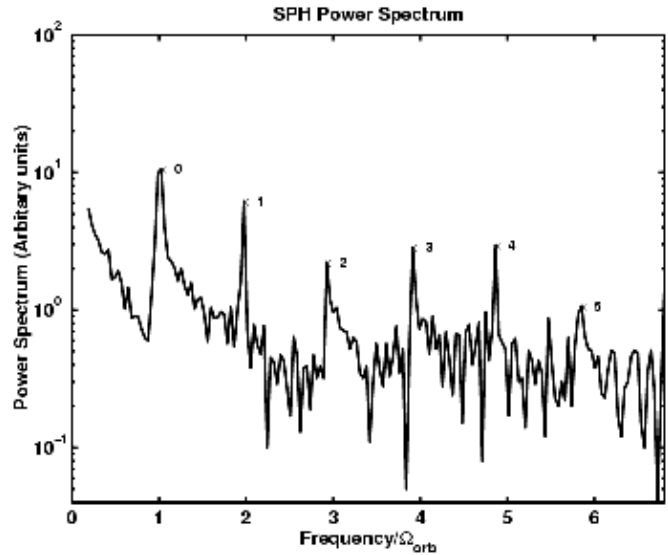,width=0.5\textwidth,angle=0}
  \caption{
           Power spectrum of the SPH dissipation light curve Figure \ref{figure:dissipation_plot} (a). The time series covered approximately 60 superhump periods with a resolution of $0.01\Omega_{orb}^{-1}$. The peak labelled 0 is the superhump frequency, the superhump harmonics are also labelled (1-5). The lower peaks correspond to linear combinations of $\Omega_{orb}$, $\omega$ and $\Omega$.
          }
  \label{figure:power_spectrum}
\end{figure}

To determine the superhump period, $P_{sh}$, we took 
a Fourier transform of approximately sixty superhump cycles of the simulated light 
curve. The power spectrum is shown in Figure 
\ref{figure:power_spectrum}. The spectrum is plotted using a 
logarithmic scale to reveal the lower level peaks. A large peak that corresponds to the superhump period, labelled 0 in the figure, dominates the spectrum; the superhump harmonics have also been labelled (1-5). Using the power spectrum we estimated the superhump period to be $(1+0.0222\pm0.0005)P_{orb}$; 
this implies $P_{prec} = (46 \pm 1) P_{orb}$.
The superhump period manifest in our simulation is in agreement
with the analysis of 
Mineshige, Hirose \& Osaki (1992).
%\cite{Mineshige:1992}.
The other much lower amplitude 
periodicities present occur at frequencies $\Omega$ 
related to $\Omega_{orb}$ and disc apsidal precession frequency $\omega$
by:
$$k(\Omega-\omega) = j(\Omega-\Omega_{orb})$$
\noindent where k, j are positive integers. 

\subsection{Trailed Spectra}

Kinematics in observed systems are revealed primarily by the spectral line profiles. To
compare the results of the two simulations with CV and SXT observations we therefore generated
synthetic emission line profiles and used these to build up synthetic trailed spectrograms.

An accretion disc emission line is Doppler broadened. Horne \& Marsh (1986) calculated
spectral line profiles from discs with smooth azimuthally symmetric line emissivity
distributions, and Horne (1995) further included the effects of anisotropic turbulence on the
local line broadening profile. We generated artificial emission line spectra for emissivity
distributions calculated from both the simple 3D code and the 2D SPH code.  

\begin{figure*}
  \psfig{file=\figdir/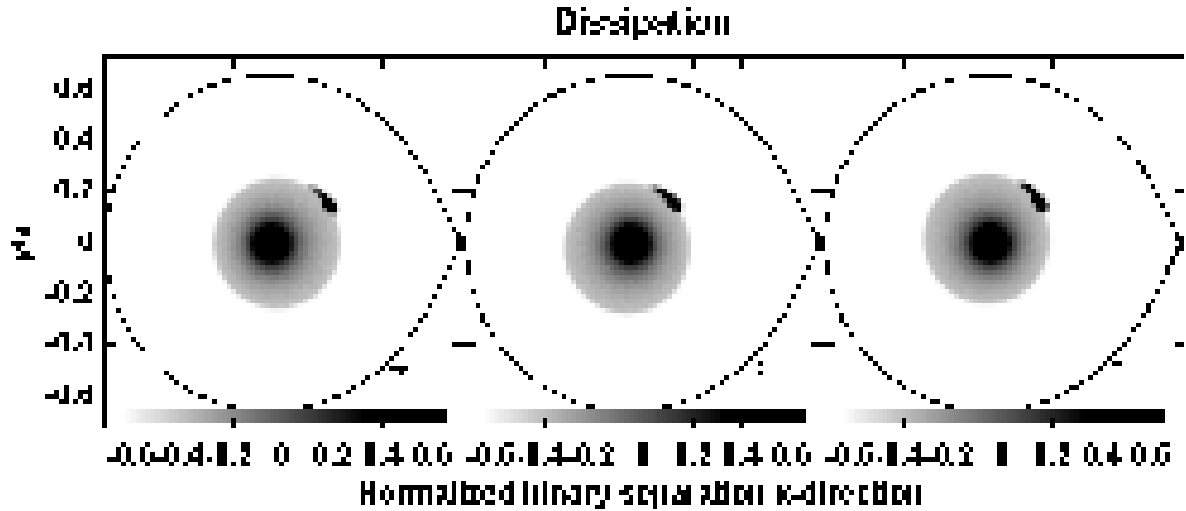,width=1.0\textwidth,angle=0}
  \caption{
Dissipation maps for the accretion disc used in the analytical model. The dissipation is diplayed using a logarithmic greyscale for binary phases 0.00, 0.36, and 0.64 respectively. 
The solid curve is the Roche lobe of the primary. 
The secondary's orbital motion is vertially upwards. 
The arrow displayed in each map indicates a fixed direction to an
observer, with phases defined as in Figure \ref{figure:density_maps}. 
          }
  \label{figure:3d_xy_plot}
\end{figure*}

\begin{figure*}
  \psfig{file=\figdir/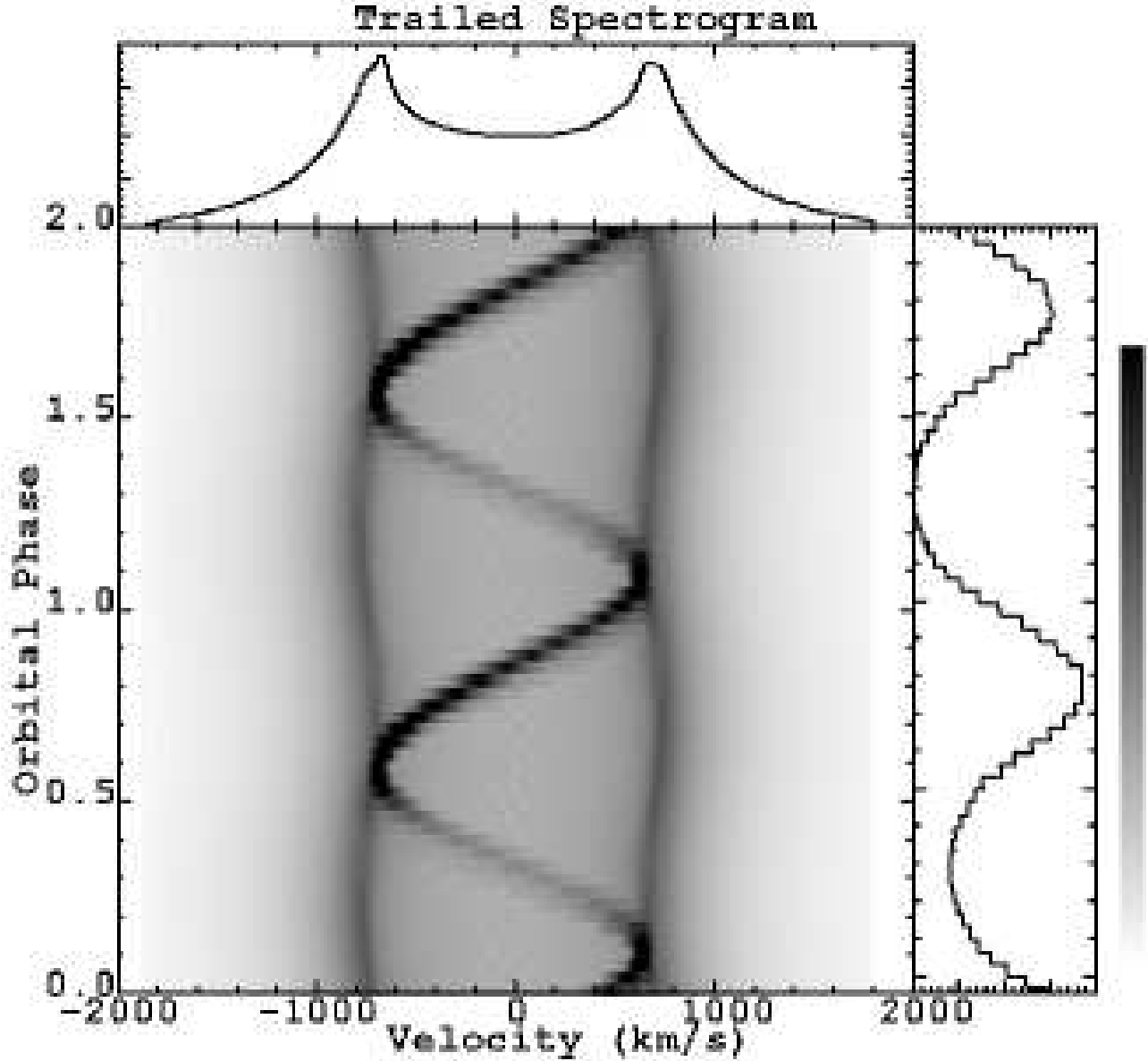,width=1.0\textwidth,angle=0}
  \caption{
Trailed spectrogram using a linear greyscale for the analytical accretion disc model. The spectrogram is for two orbital periods. The curves on the top and right of the spectrogram are the average line flux vertically and horizontally respectively, i.e. they constitute a mean line profile and a light curve respectively.  
          }
  \label{figure:3d_2orbs_trailed_spectra}
\end{figure*}

\begin{figure*}
  \psfig{file=\figdir/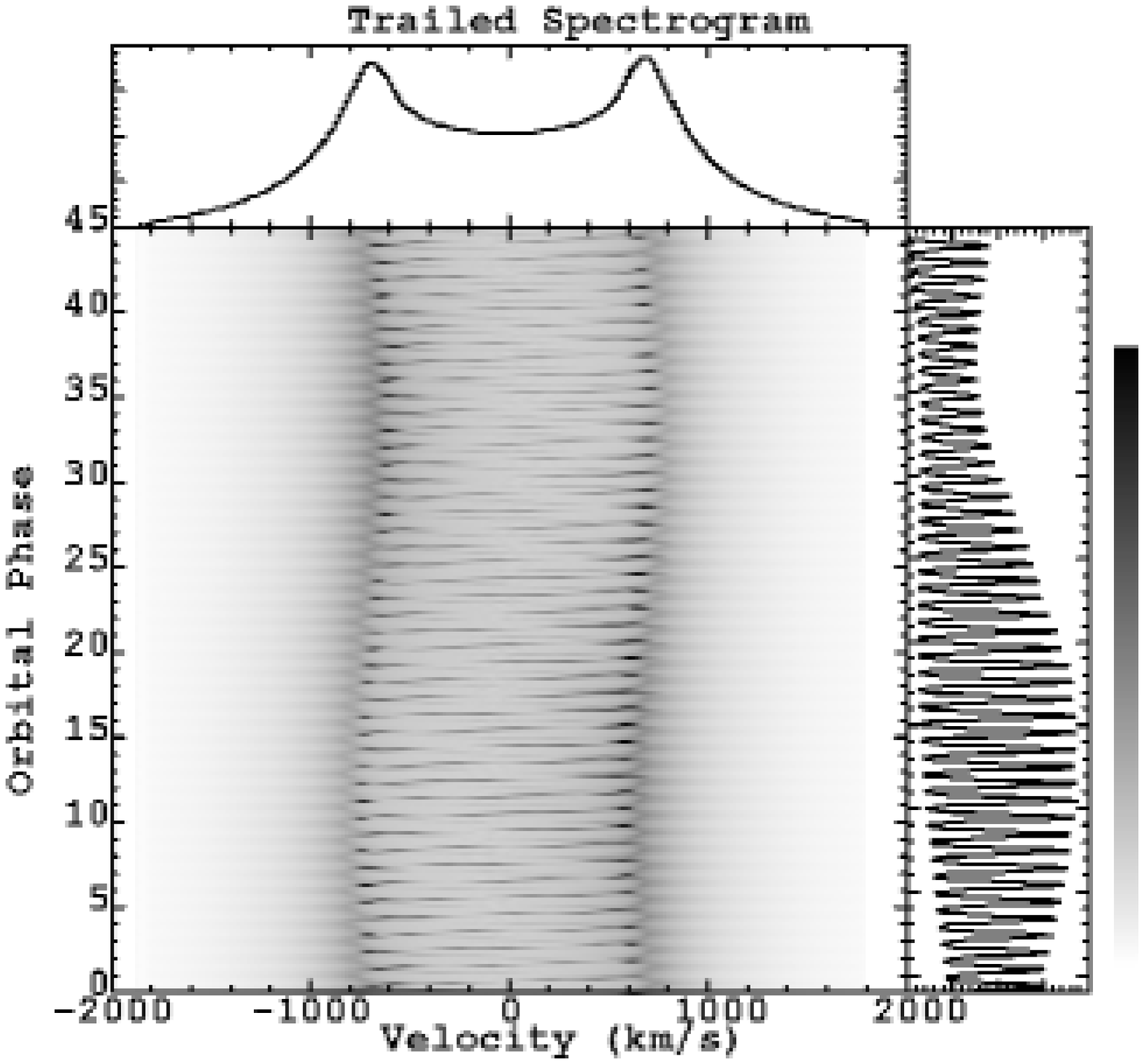,width=1.0\textwidth,angle=0}
  \caption{
Trailed spectrogram using a linear greyscale for the simple 3D accretion disc model. The spectrogram is for a complete disc precession. The curves on the top and right of the spectrogram are the average flux vertically and horizontally respectively.  
          }
  \label{figure:3d_trailed_spectra_full_prec}
\end{figure*}

For the simple 3D model we adopted Gaussian local line profiles with widths proportional to
the disc thermal velocity; the brightspot emission  was assumed to be thermally broadened
using a brightspot temperature of 10$^4$K. We did not check for occultation of the accretion
disc by the brightspot, but the $70^{\circ}$ inclination was chosen to avoid eclipses by the
secondary.  Figure \ref{figure:3d_xy_plot} shows three dissipation maps from the analytical
model for orbital phases 0.00, 0.36 and 0.64 which correspond to the same orbital phases used
in Figure \ref{figure:density_maps}. These maps show the dissipation plotted using a
logarithmic greyscale with white low dissipation and black high. The primary Roche lobe has
been plotted using a solid curve. The arrows indicate a fixed direction, as
in Figure~\ref{figure:density_maps}. The extent of the
hotspot region is clearly visible in these maps. The simulated trailed spectra for two
orbital periods is shown in Figure \ref{figure:3d_2orbs_trailed_spectra} and the trailed
spectra for a full disc precession cycle is shown in Figure
\ref{figure:3d_trailed_spectra_full_prec}. 

\begin{figure*}
  \psfig{file=\figdir/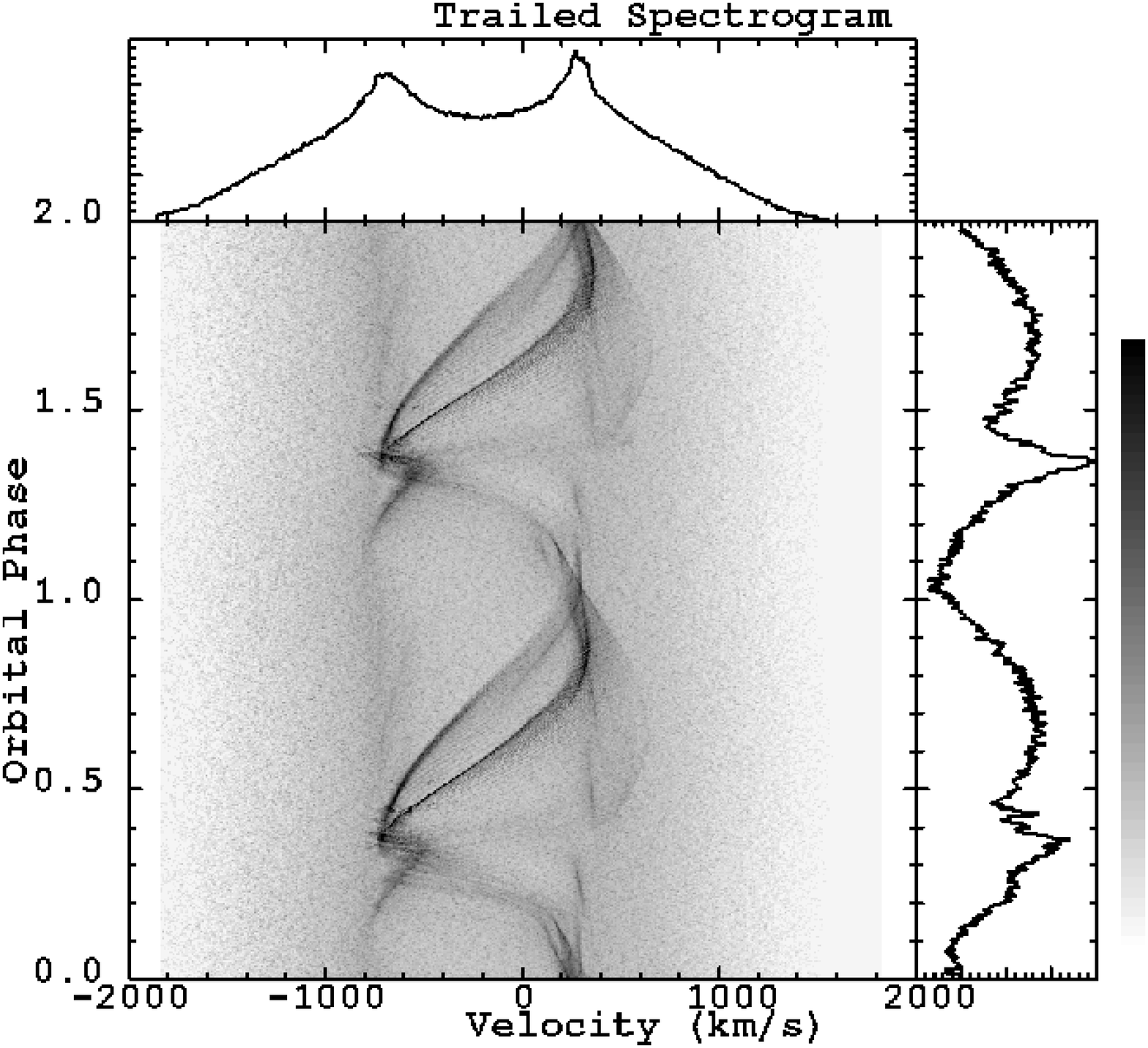,width=1.0\textwidth,angle=0}
  \caption{
Trailed spectrogram using a linear greyscale for the SPH accretion disc model. See Figure \ref{figure:3d_2orbs_trailed_spectra} caption.
          }
  \label{figure:sph_trailed_spectra}
\end{figure*}

\begin{figure*}
  \psfig{file=\figdir/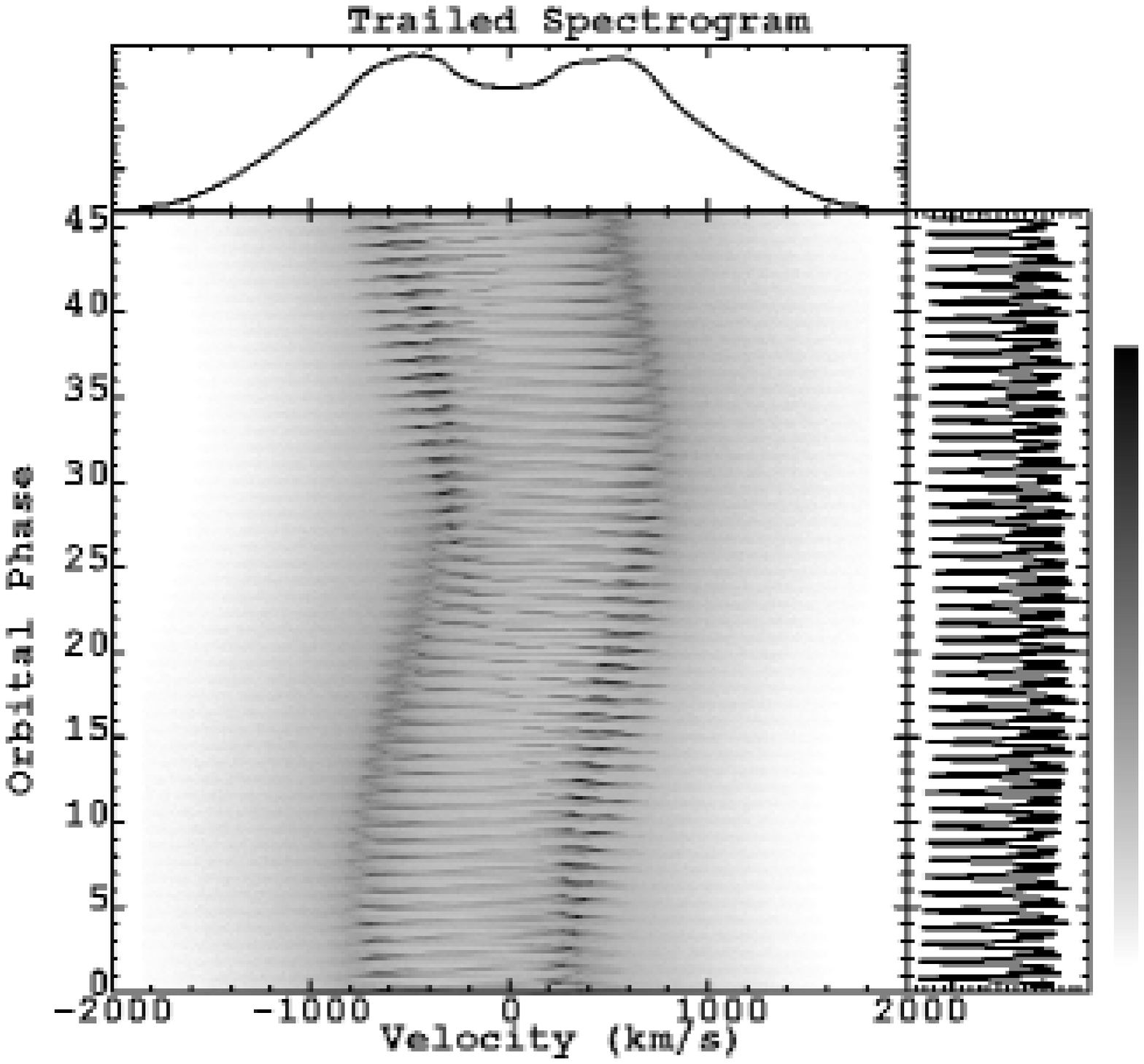,width=1.0\textwidth,angle=0}
  \caption{
 Trailed spectrogram using a linear greyscale for the SPH accretion disc model. The spectrogram is for a complete disc precession. The curves on the top and right of the spectrogram are the average flux vertically and horizontally respectively. 
}
  \label{figure:sph_trailed_spectra_complete_prec}
\end{figure*}

Using the SPH model the dissipation from the accretion disc is divided into 
radial velocity bins. The emission in each velocity bin 
is the sum of the viscous dissipation from each particle 
within the velocity bin. For each SPH time step, 
$0.01\Omega_{orb}^{-1}$, an emission line profile was 
generated and shown in the trailed spectrogram in
Figure \ref{figure:sph_trailed_spectra}. We have used a linear greyscale to represent dissipation intensity. 
The bottom of the spectrogram corresponds to the minimum flux 
point on the artificial light curve of Figure \ref{figure:dissipation_plot}
(labelled `$min$'). 
At approximately orbital phase 0.36 (`$max1$') there is a large emission associated with the first 
large spiky hump on the light curve. At about 
orbital phase 0.64 (`$max2$') there are two distinctive 
emission arcs. These arcs are generated by the particles in the brightspot region. The lower arc is from 
the particles in the gas stream that collide with the 
particles in the accretion disc. The other arc is 
the emission from the particles in the accretion disc that 
are in the brightspot region. 
The disc spiral density compression waves produce enhanced emission
that can be seen
in Figure~\ref{figure:sph_trailed_spectra}
 as features in the trailed spectrogram along
the velocity lines $\sim$$300$ \kms and $\sim$$-700$ \kms.
Figure 
\ref{figure:sph_trailed_spectra_complete_prec} 
shows the trailed spectrogram for a complete disc precession cycle,
equivalent to that of the simple 3D model in
Figure \ref{figure:3d_trailed_spectra_full_prec}.

All of our simulated trailed spectrograms are dominated by strong double-peaked emission
which exhibits significant variability. Emission from the brightspot region is prominent in
all spectrograms. In Figures \ref{figure:3d_2orbs_trailed_spectra} to 
\ref{figure:sph_trailed_spectra_complete_prec} the summation of the emission horizontally and
vertically is shown, so the upper panel is the mean line profile, and the right-hand panel is
the light curve. The light curve in Figure \ref{figure:3d_trailed_spectra_full_prec} clearly
shows the modulation in the superhump profile which was fully explored in RHP2001 and Rolfe
(2001).

\begin{figure*}
  \psfig{file=\figdir/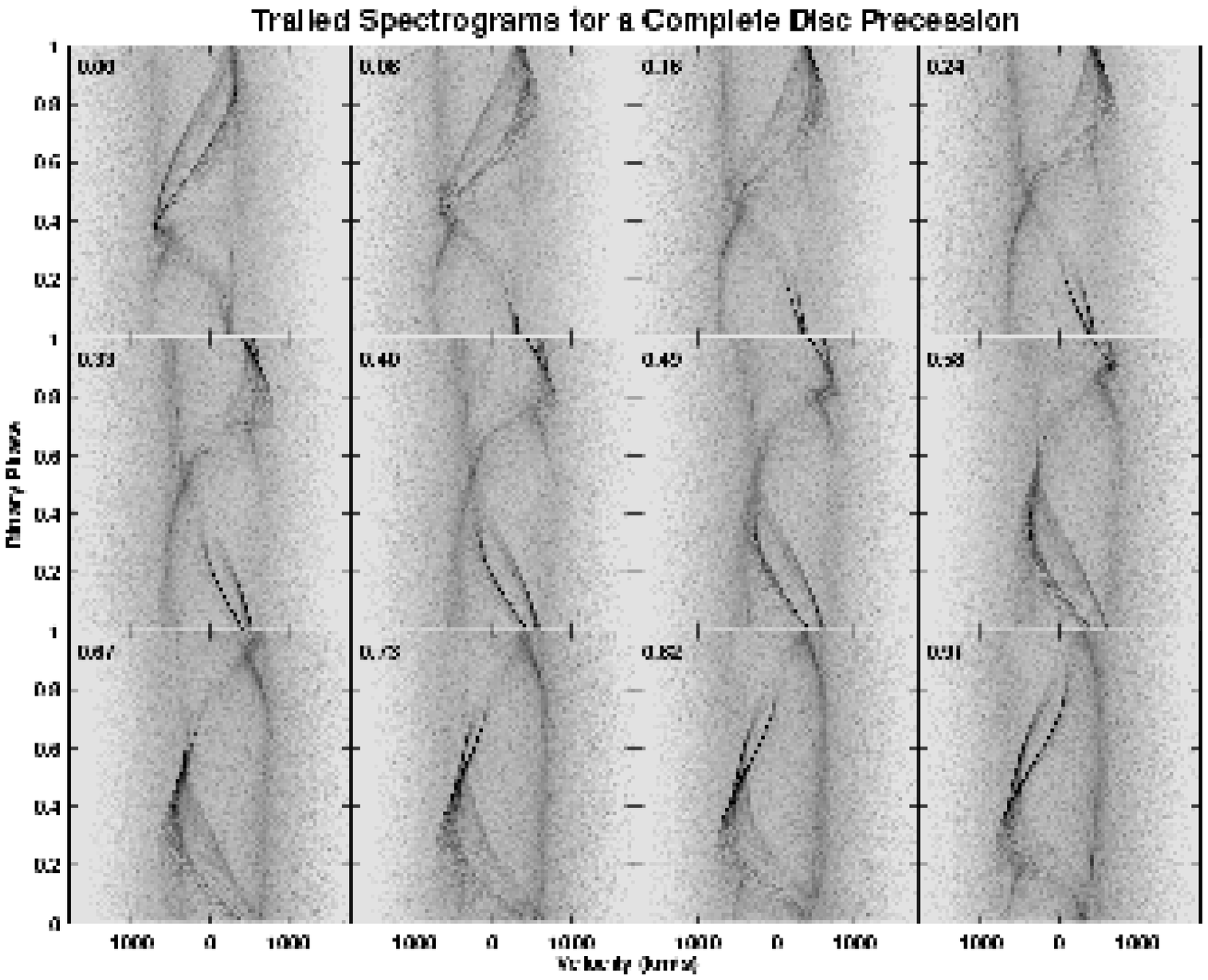,width=1.0\textwidth,angle=0}
  \caption{
          Trailed spectrograms for a complete disc precession from the SPH simulation. The spectrograms use a linear greyscale and the number in the upper left-hand corner of each image indicates the disc precession phase.  
          }
  \label{figure:total_spec}
\end{figure*}

\begin{figure*}
  \psfig{file=\figdir/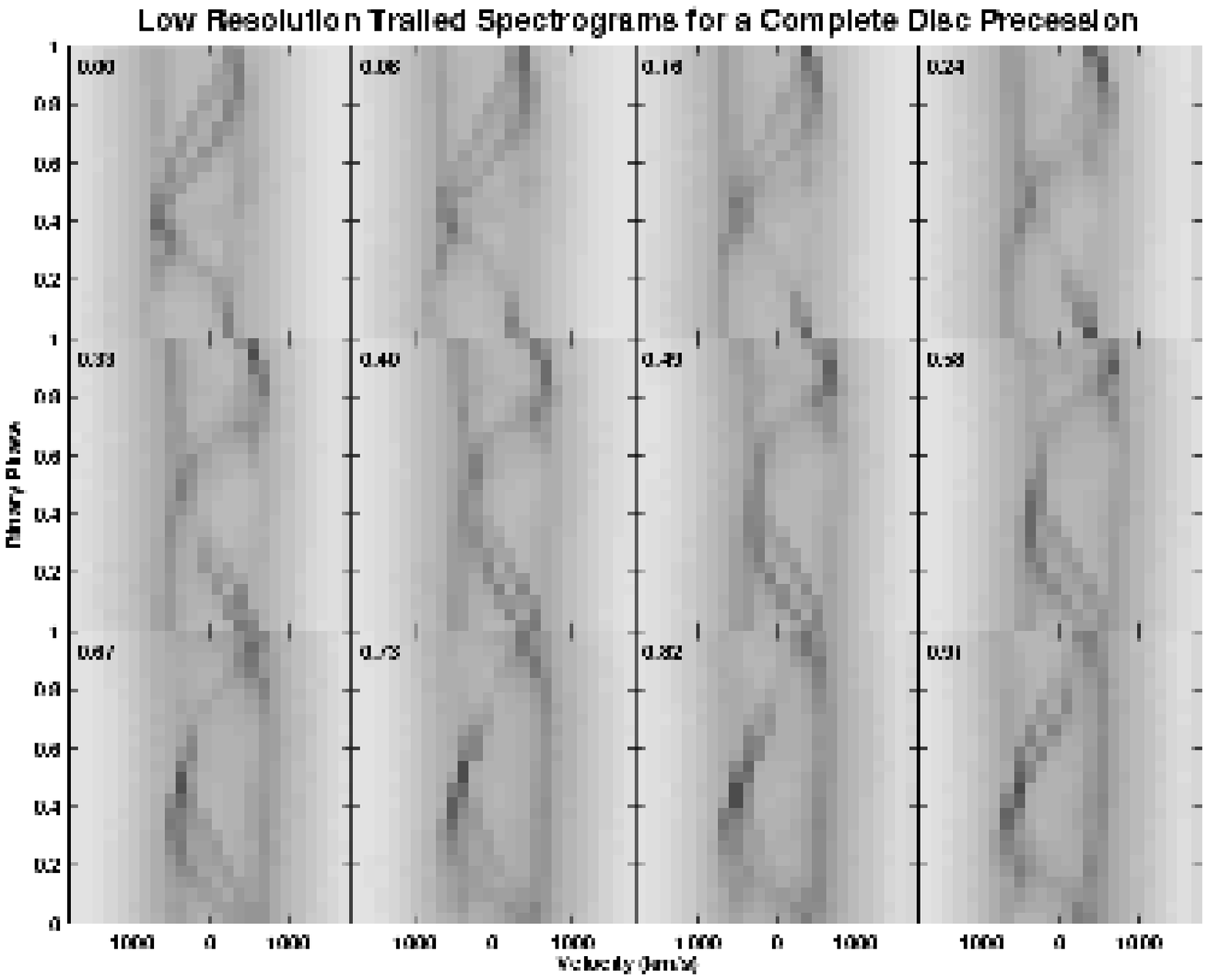,width=1.0\textwidth,angle=0}
  \caption{
          Low-resolution trailed spectrograms for a complete disc precession from the SPH simulation. The spectrograms use a linear greyscale and the number in the upper left-hand corner of each image indicates the disc precession phase.  
          }
  \label{figure:low_res_total_spec}
\end{figure*}

Figure \ref{figure:total_spec} is a montage of SPH trailed spectra for different orbits
sampling a complete disc precession. Each spectrogram uses the same linear greyscale and the
number in the upper left-hand corner of each image indicates the disc precession phase. This
figure clearly demonstrates the variability of accretion disc spectra with respect to disc
precession phase. We have placed on the Internet a Microsoft AVI movie 
showing the spectrogram's evolution over a complete disc precession cycle 
\cite{URL:Spectrogram}.   In order to compare our
high resolution trailed spectra with current observations, we reduced the spectral
and time
resolution of each spectrogram of Figure \ref{figure:total_spec} by a factor of 25 and the
result is displayed in Figure \ref{figure:low_res_total_spec}. 

In all of our trailed spectrograms we see the effects of the precessing disc.
In Figure~\ref{figure:3d_trailed_spectra_full_prec} we see a long period S-wave in the envelope of the disc emission, caused by the disc's precession on $P_{prec}$. The orbital modulation in the brightspot emission causes the 
shorter timescale S-wave which is most clearly apparent where it crosses each of the two peaks in the line profile.
%The slow evolution in brightness of the two peaks in the line profile
%with the red-shifted peak brightening to a maximum at $\phi_{orb} \sim12$ then fading, while the blue-shifted maximum reaches a peak at 
%$\phi_{orb} \sim33$ is due to the movement (in inertial
%space) of the location on the disc rim at which the dissipation
%reaches a maximum.

In Figure~\ref{figure:sph_trailed_spectra} we show the more complex 
orbital modulation in the line profile which occurs in the SPH simulation.
At first glance the pattern appears to repeat from orbit to orbit,
but in fact there is a gradual evolution which can be clearly seen
in Figures \ref{figure:sph_trailed_spectra_complete_prec} and \ref{figure:total_spec}. The spectral pattern repeats after a complete disc precession period.
 
\subsection{Doppler tomography}

Doppler tomography was developed to interpret trailed spectrograms of CVs \cite{HorneMarsh:1988,Marsh:2000}. Generally observers use a maximum-entropy inversion process to generate the most likely Doppler map from the spectra. Doppler tomography condenses the information in a trailed spectrogram into a velocity map of the emission regions in a system. The observed spectral line profiles as a function of orbital phase are inverted to reveal the best-fitting Doppler map of the line flux as a function of velocity in the binary frame. 
It implicitly assumes:
\begin{itemize}
\item
the emission regions are fixed within the co-rotating frame 
\item
the emission does not vary with time over the course of a binary orbit
\item
all points are visible at all times
\item
all motion is in the orbital plane
\item
the width of the profiles from any point is small
\end{itemize}

Our simulations, those described in \cite{Murray:2000} and eclipse mapping of V348 Pup
\cite{RolfeEt:2000} reveal discs which violate the first two of these assumptions. Hence,
for precessing non-axisymmetric discs, it is safer to work with trailed spectrograms
directly rather than constructing Doppler tomograms. Nevertheless, many  of the relevant
observational  publications in the last decade have included tomography,  so we have
constructed tomograms from our simulated line profiles.

\begin{figure}
  \psfig{file=\figdir/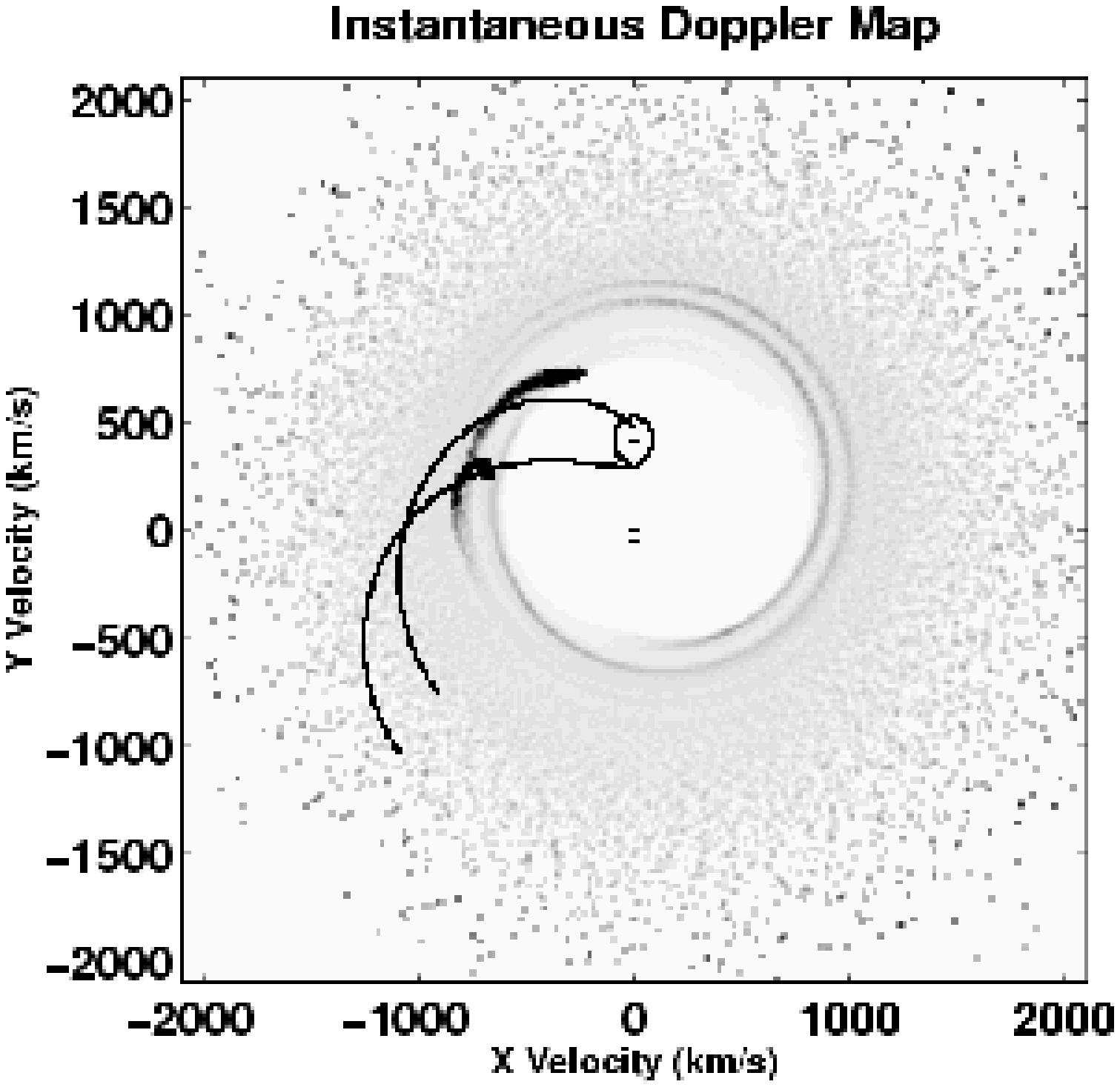,width=0.5\textwidth,angle=0}
  \caption{
           An instantaneous SPH model Doppler map. The top small black cross is the velocity co-ordinate of the secondary. The other two small black crosses are the velocity co-ordinates of the centre of mass of the binary system and the primary respectively. The predicted velocities of the gas stream are plotted and the velocities of the accretion along the gas stream. The velocity projection of the secondary's Roche lobe is also plotted. 
          }
  \label{figure:doppler_map_one_time_step}
\end{figure}

Doppler maps represent the binary configuration in two-dimensional velocity-space. 
Figure~\ref{figure:doppler_map_one_time_step} shows  an example of an
instantaneous Doppler map generated from the SPH dissipation data.  The primary
and secondary velocities are overlaid on the Doppler map y-axis  with the
secondary having a positive y-velocity.  The lower of the two overplotted arcs
indicates  the ballistic trajectory of the gas stream; the upper  arc indicates
the velocity of a circular Keplerian disc at positions along the ballistic stream
trajectory: the stream's `Kepler shadow'. The velocity of each particle is used to transformed each particle into velocity space. The velocity density at each velocity cell in the Doppler map is the sum of the dissipation from each particle mapped to that cell, i.e. the number of particles occupying each velocity cell multiplied by the particle dissipation, is plotted using a linear grey intensity scale. 

Figure \ref{figure:density_maps}($d$, $e$, $f$) demonstrates that
the emission from the disc changes substantially in the course 
of a binary orbit, and is far from fixed in the co-rotating frame.
Hence the instantaneous 
Doppler map varies significantly with orbital phase. 
Figure \ref{figure:density_maps}($g$, $h$, $i$)
shows detail of instantaneous Doppler maps corresponding to
the dissipation snapshots in Figure \ref{figure:density_maps}($d$, $e$, $f$).

The prominent dissipation near the letter $``G"$ 
on Figure \ref{figure:density_maps}($i$) 
arises from the gas stream as it impacts the outer edge of the accretion disc. 
This gas is accelerated until it has the same speed as the gas in the disc,
producing the small very bright arc close to the impact point. 
The letter $``H"$ on this Doppler map indicates the arc of emission produced
by the disc particles in the impact region.
As the two flows converge (progressing to velocity locations which 
move in an anti-clockwise sense) the relative
velocity of the two arcs decreases to zero. 
Emission from the gas downstream of the impact point can be 
seen trailing away (anti-clockwise) from the brightspot impact point. 
The outer disc in the simulation
does {\bf not} consist of particles executing circular Keplerian orbits, so
the disc velocity at the impact point does {\bf not} generally lie on
the Kepler shadow trajectory overplotted on the map. 

In Figure~\ref{figure:density_maps}($g$) there are {\it three} arcs of enhanced emission. 
Two of them are produced by the stream particles and the disc particles at the impact; the
third, uppermost, velocity arc is produced by emission from the centrifugally expelled
material, which can be seen extending outside the primary Roche lobe in
Figure~\ref{figure:density_maps}($d$). The material outside the lobe is moving up
and slowly to the right in Figure~\ref{figure:density_maps}($d$), and so it appears in the upper
right quadrant of the velocity map in Figure~\ref{figure:density_maps}($g$), where it extends
across the donor velocity lobe. 

The disc in our simulation is eccentric, precessing rapidly 
with respect to the co-rotating frame (since it is precessing
only slowly in the inertial frame),
and continuously flexing and relaxing on the superhump period.
This causes the
appreciable
changes in the instantaneous
Doppler maps produced at different orbital phases, Figure \ref{figure:density_maps}($g$, $h$ and $i$).   

\begin{figure}
  \psfig{file=\figdir/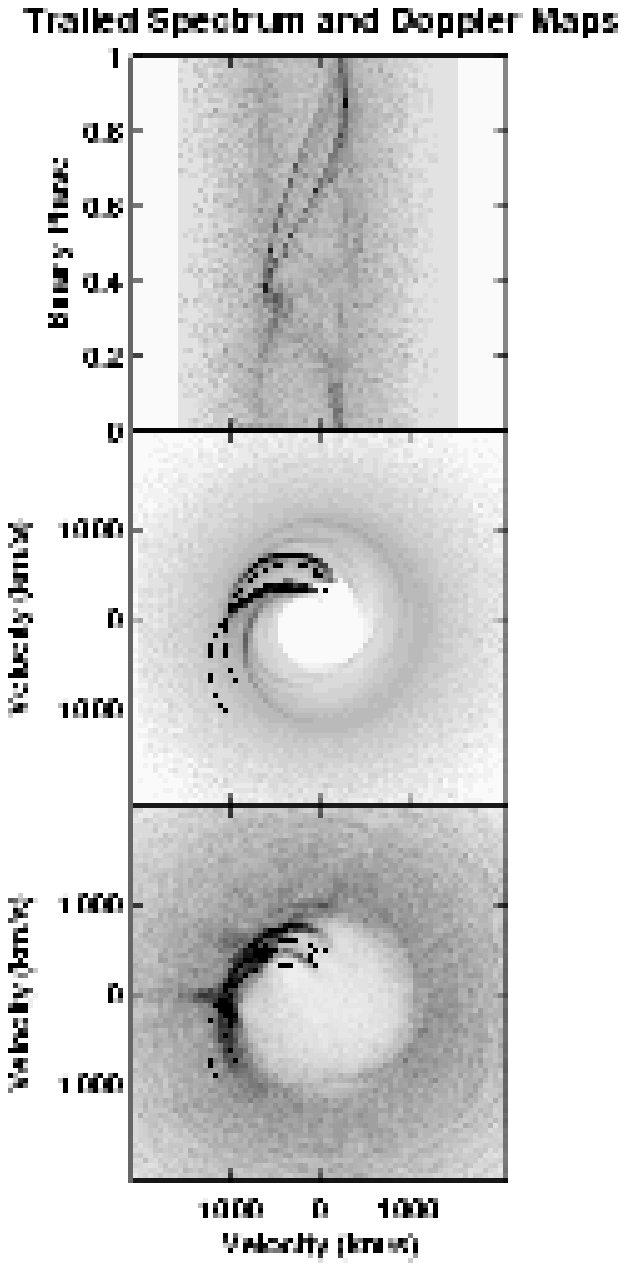,width=0.5\textwidth,angle=0}
  \caption{
           The top image is a SPH simulation trailed spectrogram
for a single orbit. 
The middle image is the corresponding direct
SPH model Doppler map, which was calculated 
by summing the instantaneous
Doppler maps for each timestep in the orbit.
The high intensity emission arc near the 
stream ballistic path was generated by the particles 
in the gas stream colliding with the outer edge of the 
accretion disc. The upper bright arc was generated by the disc
particles in the impact region. 
The bottom image is the corresponding
maximum entropy
Doppler tomogram.  
          }
  \label{figure:doppler_map_average}
\end{figure}

Since observers cannot produce an instantaneous Doppler map,
we show in
Figure \ref{figure:doppler_map_average}, middle image, a Doppler dissipation 
map for a complete orbital period. This map was generated by averaging 
the instantaneous Doppler maps for each model time step, $0.01\Omega_{orb}^{-1}$, i.e. 629 ($\approx 100 \times 2 \pi$) instantaneous maps corresponding to a complete
orbit were averaged. 
Two large emission arcs dominate this map. 
The lower arc is situated just above the gas stream 
ballistic trajectory
and was generated by the gas stream impacting the outer edge of the 
accretion disc. 
The merging streams generate a large dissipation.
The upper arc is 
generated by the particles in the accretion disc that interact 
with the gas stream particles. 
These two bright arcs show the change of velocity (and position) 
of the impact region as a function of orbital phase. 
There is also another lower intensity arc in the lower left-hand quadrant. The source of this arc is one of the spiral density
waves seen in Figure  \ref{figure:density_maps}($d$).

\begin{figure}
  \psfig{file=\figdir/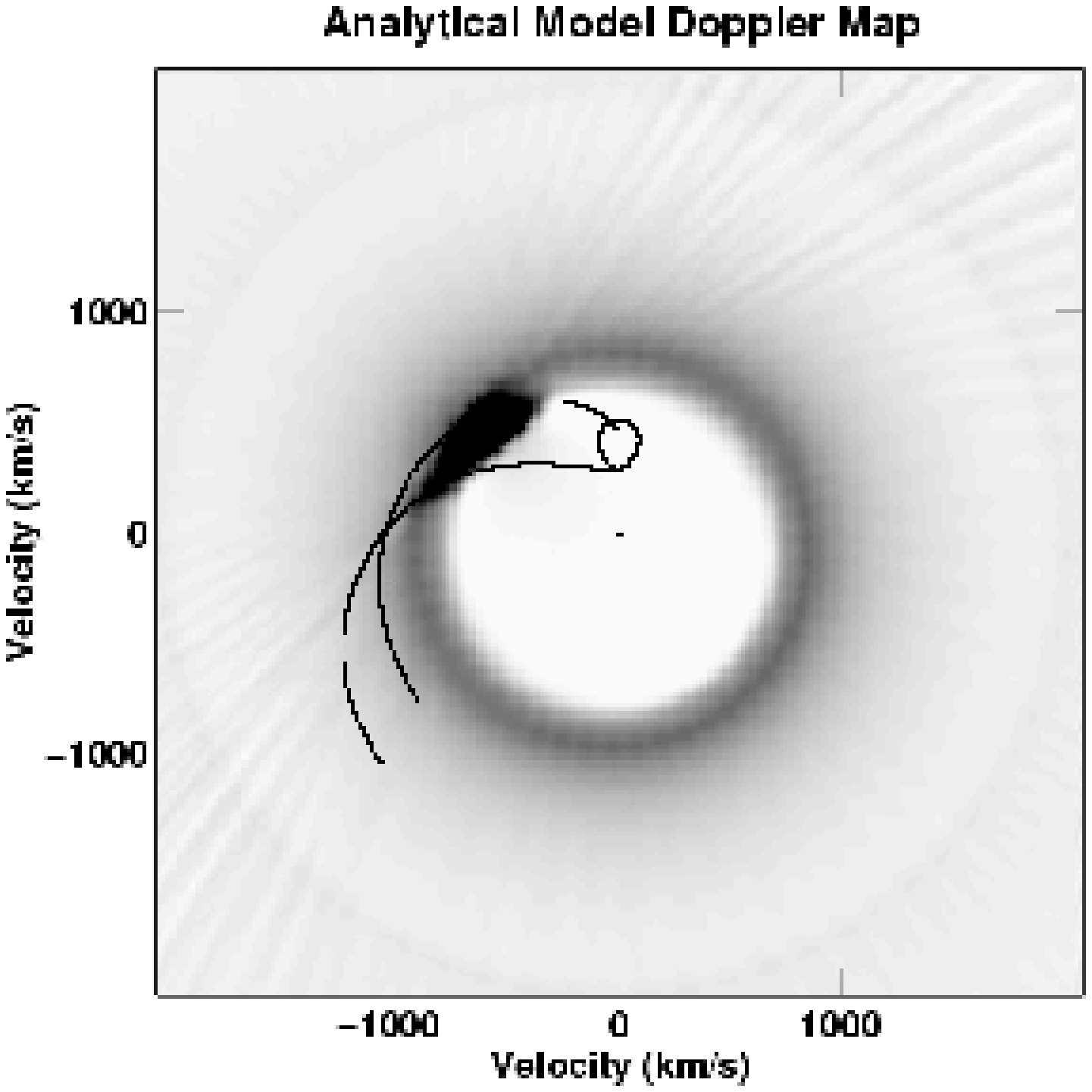,width=0.5\textwidth,angle=0}
  \caption{
           Analytical model Doppler map. See caption of Figure \ref{figure:doppler_map_one_time_step} 
          }
  \label{figure:doppler_map_dans}
\end{figure}

The SPH trailed spectrogram, shown in  Figure \ref{figure:doppler_map_average},  was inverted
using {\sc doppler}, a maximum entropy Doppler tomography  package developed by Tom Marsh
\footnote{http://www.astro.soton.ac.uk/$\sim$trm/tar/doppler.tar.gz},  and is shown in the 
bottom image of Figure \ref{figure:doppler_map_average}.  A constant Doppler map was generated
using {\sc makimg} and then scaled using {\sc optscl}.  The Doppler map had a starting
chi-squared value of 26.6 and was slowly reduced to 10.0 using {\sc memit}. The resulting maximum
entropy Doppler map has a much reduced resolution compared to the  Doppler map generated 
directly from the SPH simulation particle data (shown in the middle image of 
Figure~\ref{figure:doppler_map_average}) and the disc spiral density waves are now poorly
defined. The inversion process has reproduced  two high intensity emission arcs located near  the
gas stream ballistic path and the gas stream Kepler shadow, but their appearance differs from
those in the direct SPH map.  A number of artifacts are apparent in the maximum entropy map. 

Figure \ref{figure:doppler_map_dans} shows a Doppler map using the velocity data from the 3D analytical model over a  full orbital period.  
The Doppler map was generated using the filtered back-projection  routine in 
the {\sc molly}\footnote{http://www.warwick.ac.uk/staff/T.R.Marsh/molly.tar.gz}
spectral analysis software. This Doppler map is dominated by a large emission arc associated with the dissipation from the hotspot region. The location of this emission arc lies between the gas stream ballistic path and the gas stream Kepler shadow.

Our SPH tomograms, Figures~\ref{figure:density_maps},
~\ref{figure:doppler_map_one_time_step}
and \ref{figure:doppler_map_average},
show that the emission from
the stream-disc impact can appear outside the region
between the ballistic stream
trajectory and its Kepler shadow. 
In particular, the strongest emission from the impact
(`H' in Figure~\ref{figure:density_maps}(i))
is outside this region.
Hence tomographic analysis such 
as the determination
of $K_{2}$ for AM CVn by Nelemans, Steeghs \& Groot (2001)
may be flawed.
%1,8,9

\section{Discussion \& Conclusions}
\label{sec:Conclusions}  

We have presented two simulations of a binary system with an extreme mass
ratio of $q=0.1$. Using a SPH simulation of the accretion disc we  find the disc
to be eccentric, non-axisymmetric  and precessing in the inertial frame.

The light curve in our SPH simulation shows complex variations  with a short
duration  spiky peak superimposed on a smoother modulation. The light curve 
roughly repeats  on the superhump period, but evolves continuously from cycle
to cycle, as the example of two consecutive  orbits shown in 
Figure~\ref{figure:sph_trailed_spectra} shows. The power spectrum of the
simulated light curve is dominated by  the superhump frequency and its
harmonics. 

Simulated SPH trailed spectrograms show complex fine structure that changes  from
one orbit to the next. A full disc precession cycle elapses  before the trailed
spectrogram repeats. The trailed spectra have  changing features that include
two very bright emission arcs associated  with the impact of the gas from the
donor star  on the outer edge of the accretion disc. The 3D model trailed
spectrograms show less complexity than the SPH model. These spectrograms show bright `S' wave features that slowly change as the elliptical disc precesses in the binary frame.  The simulated dissipation light curve from this model is also modulated due to the precession of the accretion disc.

The differences in the trailed spectrograms arise from the differences
in the two models: the 3D model
underlying Figure~\ref{figure:3d_trailed_spectra_full_prec} smoothly distributes
weight across all the mass in the disc.   Its precursor 
was generated to explain the skewed absorption lines produced by AM CVn's
precessing  disc \cite{PHS:1993}.  In contrast, the SPH trailed spectrogram 
(Figure~\ref{figure:sph_trailed_spectra_complete_prec}) is weighted by
dissipation. Consequently it reveals those locations where the flow is strongly
sheared, or subject to compression under the tidally-modulated flexing.  These
regions will contribute disproportionately to emission line production. Hence,
treating the disc component as in section 2.2 may be more appropriate for
comparison with absorption lines from precessing discs, while the approach of
section 2.3 is more appropriate for comparison with emission lines produced by
such discs.  

Only 8 CVs and 8 LMXBs of $\sim 600$ systems listed in Ritter and Kolb (2003)
have measured mass ratios $q\leq 0.125$. 
To our knowledge none of these have been 
observed  extensively enough to produce trailed spectrograms uniformly
sampling the 
disc precession cycle. The best observations to compare
with our simulations are those of XTE\thinspace J1118+480 in its 2000
outburst. Torres et al. (2002) show trailed spectrograms taken at four epochs
during the outburst, from which they generate Doppler tomograms. At all
epochs these tomograms are dominated by emission from the stream-disc impact
region.  The spectrograms are of considerably lower signal-to-noise ratio
than those shown in Figure~\ref{figure:low_res_total_spec} but it is clear that
the brightest emission moves from the red side 
(at orbital phase $\sim 1$; c.f. disc precession phase 0.24 in our
SPH trails) to the blue side (at orbital phase $\sim 0.5$;
c.f. disc precession phase 0.73 in our SPH trails)
over an interval of $\sim$20 days. It is clear that the observed
S-wave is non-sinusoidal, and its shape changes.  
These characteristics are all seen in our simulations.
Further, higher resolution, observations of XTE\thinspace J1118+480
(Haswell et al 2004) show in more detail that the observed S-waves
exhibit shapes and phasing consistent with our simulations.
VW Hyi, a CV with $q \approx 0.17$ (Schoembs \& Vogt 1981),
shows hints of two distinct S-waves in the trailed spectrogram
presented in Figure~A.14 of Tappert et al (2003), though the
signal-to-noise ratio is again rather lower than in our simulated
trails.
We conclude our SPH simulations do appear to
reproduce at least some observed characteristics
of some extreme mass ratio systems.

Instantaneous Doppler maps generated from the SPH simulation show  bright
arcs created by the impact of the gas stream  with the outer edge of the accretion
disc. As the gas stream interacts with the disc material much
energy is   dissipated. Integration of the SPH instantaneous Doppler maps over a full binary orbit shows two large very bright emission arcs resulting from the changing shape of the accretion over a binary period. The arcs  indicate the changing velocities at
the impact point. The 3D model Doppler map shows a single large emission arc associated with the brightspot region and is located between the gas stream and the stream's `Kepler shadow', see Figure \ref{figure:doppler_map_dans}. 

In our simulations, the stream-disc impact region is
a prominent source of dissipation.
It is generally accepted that for CVs in superoutburst 
tidally-modulated viscous dissipation in the bulk of the disc generates 
the superhump light (e.g. Murray 2000).  
These are  known as normal or `early' superhumps.  These sometimes evolve to 
`late' superhumps as the disc returns to its low-viscosity 
quiescent state. In late superhumps 
the stream-disc impact dissipation is a prominent feature of the CV light curve 
(RHP2001). 
As we noted in section 2.3, the prominence of the stream-disc impact region may
to some extent be an artifact of our 2D simulations. However this region was
also observed to be prominent in 
the outburst trailed spectra of
XTE \thinspace J1118+480.  As XTE \thinspace J1118+480 returned
to quiescence, Zurita et al. (2002) observed trailed spectra in which 
the S-wave is much less prominent.  This observed behaviour  
is exactly the opposite of the accepted picture for CVs.
We can think of (at least) two possible reasons for this:
(i) It is possible that the fraction of the total
dissipation which occurs at the
stream disc impact region
is a function of mass ratio, becoming dominant in
outburst for very extreme mass ratio systems like that simulated herein. 
(ii) The line emission in SXTs is likely to be powered by photo-ionisation.
Thus, the observed prominence of the stream-disc impact may simply
indicate that this region is raised above the surface of the bulk of the
disc, and subtends a relatively large solid angle to the central X-ray source.

Once started, disc precession tends to persist \cite{Murray:2000}.
Hence attempts to study  SXTs upon their return to quiescence are likely to be 
complicated by phenomena related to disc precession (Haswell 1996).
For example, 
Figure~\ref{figure:sph_trailed_spectra_complete_prec}
shows that there is a radial velocity modulation on the disc
precession period.
The blue peak
of the line profile moves from $\sim$-400~\kms at around orbital phase = 30 to
$\sim$-800~\kms at around orbital phase = 10. 
An analogous, but smaller modulation is seen in
Figure~\ref{figure:3d_trailed_spectra_full_prec}.
A small part of this difference is because the SPH
trailed spectrogram corresponds to an orbital inclination of $90^{\circ}$ while
the simple 3D model uses $i = 70^{\circ}$; the SPH velocities would be reduced
by a factor 0.94 if $i = 70^{\circ}$.  The main reason for the larger velocity
modulation in the SPH simulation is the velocities of the strong
dissipation at the stream-disc
impact region vary much more than those in the simple analytic model.

Attempts have been made to measure the orbital motion of accreting  compact
objects by studying the emission lines from their accretion discs
\cite{HS90,Orosz:1994,Soria:1998}. Generally these observations are made over
several nights, sampling parts of a number of orbits, interrupted by daylight
intervals. The line profile measurements are then generally binned or folded on
orbital phase and the `orbital' motion measured by fitting a sine wave to the
radial velocities. Figures~\ref{figure:3d_trailed_spectra_full_prec} - 
\ref{figure:low_res_total_spec} demonstrate that for a precessing disc this procedure
is flawed because  the line  profiles do not repeat from orbit to orbit. 
Figure~\ref{figure:sph_trailed_spectra_complete_prec} 
shows that the radial velocity of the
outer wings of the disc emission line is as strongly modulated  on disc precession
phase as it is on orbital phase. Random, non-uniform sampling of disc precession 
phase explains why Haswell \& Shafter (1990) and Orosz et al. (1994)  found
emission line radial velocity curves which were not in phase with the compact
object's orbital motion (as inferred from the mass donor star's motion). In
contrast, for binaries where the mass ratio is not extreme enough for the disc
to extend out to the 3:1 resonance ($q \ga 0.3$) the disc is not expected to be
eccentric and precessing. For such systems it may be possible to measure the
orbital radial velocity of the compact object, if contamination of the result
by the brightspot and any other non-axisymmetric features is avoided. Indeed,
for GRO J1655-40 which has a mass ratio of $q = 0.42$ \cite{Shahbaz:2003}, 
Soria et al. (1998) found an emission line radial velocity curve which {\it is}
phased as expected for the orbital motion of the black hole.  In CVs, the
smaller dynamic range between the inner and outer disc radii may make 
measurements uncontaminated by the brightspot more challenging, nontheless
Unda-Sanzana et al. are finding encouraging results for U~Gem which has $q >
0.3$ (Unda-Sanzana \& Morales-Rueda, private communication). 

\section{Acknowledgements} SBF acknowledges the support from QinetiQ, Malvern,
in particular from David Hutber. SBF and DJR would like to thank Tom Marsh for the use
of his {\sc molly} and {\sc doppler} software. CAH  acknowledges support from
the Leverhulme Trust F/00-180/A during the early stages of this work. DJR was
supported by a PPARC studentship and a PPARC PDRA. Theoretical astrophysics
research at Leicester is supported by a PPARC rolling grant. SBF, CAH and JRM
acknowledge support from the Open University's Research School. We would also
like to thank an anonymous referee for a very detailed report and for the many
useful suggestions.

\label{lastpage}

\end{document}